\documentclass{article}
\usepackage{amsmath}
\usepackage{graphicx}
\usepackage{dcolumn}
\usepackage{bm}
\usepackage{array}
 
\newcommand\B{\rule[-2.0ex]{0pt}{0pt}}
\newcommand\T{\rule{0pt}{6.7ex}}
\newcommand\TS{\rule{0pt}{4ex}} 
\newcolumntype{M}[1]{>{\centering\arraybackslash}m{#1}}
\usepackage{hyperref}
\hypersetup{
	colorlinks=true, 
	linktoc=all,     
	linkcolor=blue,  
}
\newcommand{\sed}{scalar field energy density parameter }
\newcommand{\eosp}{effective EOS parameter }
\newcommand{\gp}{\gamma_\phi}
\newcommand{\op}{\Omega_\phi}

\begin{document}

\title{Thawing vs. Tracker Solutions: A Dynamical Systems Approach}

\author{\small Abhijit Chakraborty\footnote{ac12ms079@iiserkol.ac.in} $^{1,2}$, Narayan Banerjee\footnote{narayan@iiserkol.ac.in} $^1$, Anandamohan Ghosh\footnote{anandamohan@iiserkol.ac.in} $^1$ \\
\small $^1$Indian Institute of Science Education and Research Kolkata, Mohanpur Campus,\\ \small Nadia, West Bengal, 741246, India\\ \small $^2$University of Houston, 4800 Calhon Road, TX, 77004, USA}

\maketitle

\begin{abstract}
A comparative study of thawing and tracking models of dark energy is carried out with the help of a dynamical systems analysis. It is found that both of them have stable solutions which are consistent with the requirement of a dark energy. So none of them is actually favored from the consideration of stability. The trackers have the interesting possibility that the present acceleration is a transient phenomenon.\\

\begin{description}
\item[PACS numbers] 98.80.-k; 95.36.+x
\end{description}
\end{abstract}


\section{\label{sec:1}Introduction}

Mounting observational evidences\cite{perl, schmidt, riess, knopp} have enforced the counter-intuitive notion of an accelerated expansion of the universe. The driver of this acceleration, known as the dark energy, which produces an effective negative pressure, accounts for close to 70\% of the total energy budget of the universe\cite{planck}. Albeit its success as the possible agent driving the acceleration\cite{paddy}, the cosmological constant suffers from a huge discrepancy between the observationally required value and the theoretically predicted one\cite{weinberg}. An alternative approach is to consider an evolving dark energy where a scalar field with a positive potential, called the quintessence field has been introduced\cite{sami, varun, brax}. \\

There is a large number of scalar field models, which can quite successfully match the observational requirements but hardly any one of them has a compelling reason for its existence supported by other branches of physics. As the accelerated expansion sets in quite late in the evolution\cite{riess1}, the dark energy should start dominating over the normally gravitating matter only at a later stage. A large section of scalar field models can be characterized as thawing or freezing depending on the way it starts dominating over the other matter components that gravitate in the usual attractive manner\cite{caldwell}. A thawing model is one where the equation of state parameter $w$ is very close to $-1$ to start with, so that the energy density is nearly a constant, but increases to its present value which is still negative so as to generate a sufficient negative pressure. A freezing model, on the other hand, has an evolution at the beginning and gradually decreases to be frozen at a recent past so that $w$ attains a value 
close to $-1$. Amongst the freezing models, a class is called a tracker\cite{stein1, stein2, johri} where the dark energy density falls off almost at the same rate as dark matter (the dominating contribution of the normally gravitating matter) in the beginning. For a compact description of this classification, we refer to the work of Scherrer and Sen\cite{anjan}. \\

Both the thawing and tracking models alleviate the so-called ``cosmic coincidence'' problem: why the dark energy and dark matter are of the same order of magnitude at the present epoch\cite{concord}? This aspect of thawing and tracking models have generated a lot of attention in recent times. Aspects of a  thawing quintessence were discussed by Sen and Scherrer\cite{anjan}, Chiba\cite{chiba}, Sen {\it et al}\cite{soma} amongst others. The observational constraints on thawing models were discussed by  Chiba {\it et al}\cite{chiba2}. Tracker models are there in the literature for quite some time, some of the early references have been already mentioned. Recently there has been a thorough comparative investigation of thawing and tracker models with the aim to validate one with respect to observations\cite{shruti}.  \\

The motivation of the present work is to compare these two quintessence classes, thawing, and tracking, 
in the context of the stability criteria obtained from their respective dynamical systems \cite{glen94}. 
In a recent study the stability of some tracking models with very specific potentials was discussed \cite{nandan}
but a detailed comparative study has been lacking.
In our study we find that both the models have stable dark energy solutions. The stable situation for the thawing models are always apt to lead to an accelerated expansion in the late time, whereas for the trackers, there is an interesting possibility that the present acceleration is only a transient phenomenon, the universe will finally settle down with a decelerated expansion for all future time. However, the trackers do not rule out the possibility that the final stable configuration is that of an accelerated expansion either. 

In section 2, the Einstein field equations with a quintessence field are written in the form of an autonomous system. In section 3 and 4 we implement the conditions for thawing and tracking, respectively, and analyze the stability properties of the models. In section 5 we compare these two classes and make some concluding remarks. 

\section{\label{sec:SoE}System of Equations}
The Lagrangian corresponding to a minimally coupled scalar field is given by
\begin{equation}\label{scalar_L}
\mathcal{L_\phi} = \frac{1}{2}\partial_\mu\phi\partial^\mu\phi-V(\phi),
\end{equation} 
where $\phi$ is the scalar field and $V(\phi)$ is the potential. We consider a spatially flat, isotropic and homogeneous universe given by the metric
\begin{equation}\label{metric}
 ds^2 = dt^2 -a^{2}(t)(dr^2 + d{\theta}^2 + \sin^2 \theta d{\phi}^2),
\end{equation}
where $a=a(t)$ is the scale factor. The Friedmann equations for the universe, containing both matter in the form of a perfect fluid and the scalar field, can be written as
\begin{subequations}\label{FEH}
\begin{align}
& H^2 = \frac{1}{3}(\rho_m + \rho_\phi) \label{FEH1},\\
&\dot{H} = - \frac{1}{2}(\rho_m + \rho_\phi + \textup{p}_m + \textup{p}_\phi), \label{FEH2}
\end{align}
\end{subequations}
where $H=\frac{\dot a}{a}$ is the Hubble parameter, $\rho_m$ and $\textup{p}_m$ are the energy density and pressure of the matter distribution, respectively.
The contribution to the energy density, $\rho_\phi$, and the pressure sector, $\textup{p}_\phi$,
 are due to a scalar field $\phi$, and are given by 
\begin{eqnarray}
\textup{p}_\phi = \frac{\dot{\phi}^2}{2} - V(\phi) \label{pressure}, \\
\rho_\phi = \frac{\dot{\phi}^2}{2} + V(\phi)\label{energy}.
\end{eqnarray}
The Klein Gordon equation for the scalar field is 
\begin{equation}\label{KG}
\ddot{\phi} + 3H\dot{\phi} + \frac{\partial V}{\partial \phi} = 0.
\end{equation}
 We have used the units where $8\pi G=1$. The dark matter that fills the universe is considered to be dust, so $\textup{p}_m = 0$. It should be noted that Eq.\eqref{KG} is not an independent equation as it can be derived from the field equations \eqref{FEH} if the fluid satisfies its own conservation equation,
\begin{equation}\label{conservation}
 \dot{\rho_m} + 3H(\rho_m + p_m) = 0.
\end{equation}

We introduce a set of dimensionless variables $\op = \frac{\rho_\phi}{3H^2}$, $\gp = \frac{\textup{p}_\phi + \rho_\phi}{\rho_\phi}$ and $\lambda = - \frac{1}{V}\frac{dV}{d\phi}$. $\op$ is the \sed and $\gp = 1 + w_\phi$ where $w_\phi=\frac{\text{p}_\phi}{\rho_\phi}$ is the equation of state parameter of the scalar field.
Both these parameters, $\op$ and $\gp$ can be  estimated from observed quantities. In terms of these new variables the system of Friedmann equations can be written as
\begin{subequations}\label{diff_eq_new}
	\begin{align}
	& \Omega^\prime_\phi = 3(1 - \gamma_\phi) \Omega_\phi(1 - \Omega_\phi), \label{diff_eq_new1}\\
	& \gamma_\phi^\prime = (2-\gamma_\phi)(-3\gamma_\phi + \lambda \sqrt{3\gamma_\phi \Omega_\phi}), \label{diff_eq_new2}\\
	& \lambda^\prime = - \sqrt{3\gamma_\phi \Omega_\phi}\lambda^2(\Gamma - 1), \label{diff_eq_new3}
	\end{align}
\end{subequations} 
where $\Gamma = V \frac{\frac{d^2V}{d\phi^2}}{(\frac{dV}{d\phi})^2}$ and a prime denotes a derivative with respect to $N = \ln(a/a_0)$, $a_0$ being the present value of the scale factor.

For a scalar field used as a quintessence, the potentials can be broadly categorized into two types: thawing, and tracking (or freezing). We shall discuss them in the following two sections.

\section{\label{sec:thawing}Thawing model}
The thawing model is a quintessence model which starts with an equation of state (EOS) parameter, $w_\phi$, very close to $-1$ and slowly increases to some present value \cite{caldwell} with the evolution of the scalar field.  The corresponding potential is assumed to be a slow rolling potential which satisfies the following approximations,
\begin{eqnarray}\label{sra}
\left(\frac{1}{V}\frac{dV}{d\phi}\right)^2<<1 \label{sra1},\\
\frac{1}{V}\left(\frac{d^2V}{d\phi^2}\right)<<1 \label{sra2}.
\end{eqnarray}
The slow roll approximations were primarily used in the context of potentials responsible for the early inflation, when the scalar field energy density dominated at an early stage of evolution. For the thawing models of dark energy, the slow roll approximation does well to group together the relevant potentials.  \\

For a potential obeying slow roll approximations, we can make certain assumptions. First of all, $\gamma_\phi$ is very small since $w_\phi$ is close to $-1$. We can write $1-\gamma_\phi \approx 1$ and $2-\gamma_\phi \approx 2$ in
Eqs. \eqref{diff_eq_new1} and \eqref{diff_eq_new2}. Putting together the slow roll approximations we can write that $\lambda$ is approximately a constant, i.e.,
\begin{equation}
\lambda = \lambda_0 = -\frac{1}{V}\frac{dV}{d\phi}\Big|_{\phi = \phi_0} \label{lam_thaw},
\end{equation} 
where $\lambda_0$ is the initial value of $\lambda$, before the slow roll down the potential hill begins. We have to note here that $\lambda_0$ has to be small enough such that Eq.\eqref{sra1} is satisfied. \\

With these assumptions, the system of equations \eqref{diff_eq_new} reduces to a two-dimensional system given by,
\begin{subequations} \label{thaw_eq}
	\begin{align}
	\Omega^\prime_\phi &= 3\Omega_\phi(1 - \Omega_\phi) \label{thaw_eq1},\\
	\gamma_\phi^\prime &= -6\gamma_\phi + 2\lambda_0 \sqrt{3\gamma_\phi \Omega_\phi} . \label{thaw_eq2}
	\end{align}
\end{subequations}
The fixed points of the system of equations \eqref{thaw_eq} are given in \autoref{tab:tab1}.
\begin{table}[!h]
\caption{Fixed points for a system with a thawing potential and their stability}\label{tab:tab1}
\begin{center}
\begin{tabular}{c|c|c|c}
		Fixed point & $\Omega_\phi$ & $\gamma_\phi$ & Stability\\
		\hline A. & 0 & 0 & saddle\\
		B. & 1 & 0 & saddle\\
		C. & 1 & $\dfrac{\lambda_0^2}{3}$ \B & stable\\ \hline
	\end{tabular}
\end{center}
\end{table}

The fixed point A=(0,0) represents the beginning of the universe. At the beginning of the universe, the scale factor was zero, giving $H \rightarrow \infty$. So both $\Omega_\phi$ and $\gamma_\phi$ can be zero. To show that the fixed point is indeed an unstable one, we obtain the Jacobian of the system
\begin{equation}\label{jac_thaw}
J = \begin{pmatrix} 
3(1-2\Omega_\phi) & 0\\
\lambda_0\sqrt{\frac{3\gamma_\phi}{\Omega_\phi}} & -6 + \lambda_0\sqrt{\frac{3\Omega_\phi}{\gamma_\phi}}
\end{pmatrix}.
\end{equation}
The determinant of the Jacobian matrix, evaluated at (0,0) is negative,
indicating that the fixed point A is saddle. \\

\begin{figure}
\centering
\includegraphics[width=0.7\linewidth]{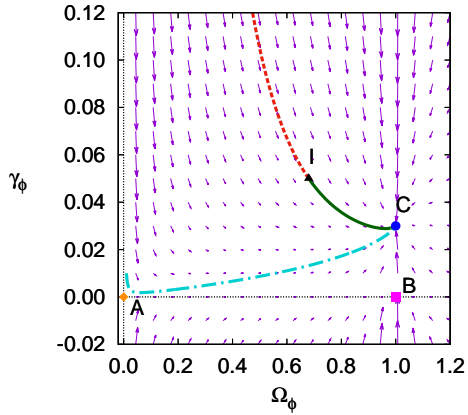}
\caption{Thawing model: Evolution of the universe in the $\gamma_\phi$ $-\hspace*{1mm}\Omega_\phi$ plane backward (red dashed line) and forward (green solid line) in time starting from the current value (I) of the scalar field energy density and effective EOS parameter. The cyan dot-dashed line shows the trajectory of the system starting from a point in the neighborhood of the saddle at A:(0,0)}
\label{fig:thawdir}
\end{figure}

The eigenvalues of the Jacobian in Eq.\eqref{jac_thaw} are $\Big\{3-6\op,-6+\lambda_0\sqrt{\frac{3\op}{\gp}}\Big\}$. For the fixed point C the eigenvalues become $\{-3,-3\}$, implying point C is a stable node since both the eigenvalues are negative. For point B, the second eigenvalue blows up to positive infinity. The other eigenvalue is negative, indicating that the fixed point (1,0) is also a saddle. To see the evolution of the universe as a system, we numerically simulate the system \eqref{thaw_eq} with initial conditions $(\Omega_{\phi_0},\gamma_{\phi_0}) \equiv (0.68,0.05)$ which are close to the currently observed values of the \sed and the \eosp\cite{ade et al}. \autoref{fig:thawdir} shows the direction field plot of the system \eqref{thaw_eq} for $\lambda_0 = 0.3$, along with the trajectory of the universe, obtained numerically. It deserves mention that the value of $\lambda$ is chosen so as to satisfy the condition \eqref{sra1}. \\

The black triangle (I) in \autoref{fig:thawdir} represents the current universe given by $(\Omega_{\phi_0},\gamma_{\phi_0}) \equiv (0.68,0.05)$. The blue dot is the point C in \autoref{tab:tab1}. The yellow diamond represents the point A (0,0) in \autoref{tab:tab1} and the magenta square is the point B (1,0). From the figure, we can infer that the universe started evolving from the saddle (0,0) and reaches the point C in future. The \sed becomes 1 at C, representing a universe completely ruled by the dark energy.\\

\begin{figure}
\centering
\includegraphics[width=0.7\linewidth]{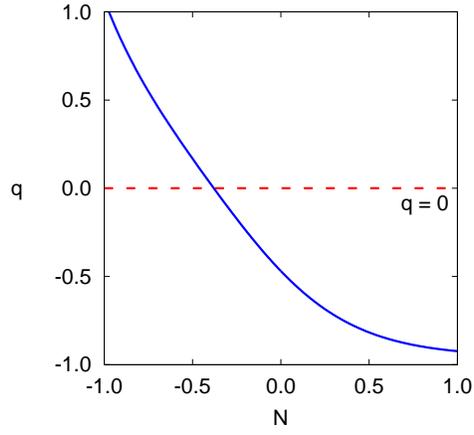}
\caption{q vs. N plot for $\lambda_0=0.3$}
\label{fig:qthaw}
\end{figure}

We now plot the deceleration parameter $q = -\dfrac{\dot{H} + H^2}{H^2}$ with $N=ln(a/a_0)$ to figure out the redshift when the current phase of acceleration begins. \autoref{fig:qthaw} indicates that the universe was expanding with a deceleration at first but as time passed, the scalar field begins to dominate, driving the universe to an accelerated expansion phase. The late time acceleration begins at $z \simeq 0.46$.\\

\begin{figure}
\centering
\includegraphics[width=0.7\linewidth]{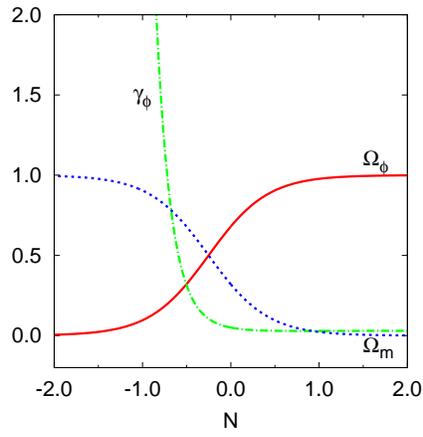}
\caption{The plot of $\op$, $\Omega_m$, and $\gp$ vs. N for a thawing dark energy}
\label{fig:thawtime} 
\end{figure}

The time series plots for $\op$, $\Omega_m$ (matter energy density), and $\gp$ are shown in \autoref{fig:thawtime}. From \autoref{fig:thawtime} we see that the scalar field energy density starts dominating over the background energy density at later stages of the evolution. It can also be seen that $\gp$ (the dot-dashed line) diverges as the system is evolved backward. This is a feature of the saddle point. A system cannot be traced back to a saddle point unless the initial condition is on the unstable manifold. Therefore it can be said that the universe starts evolving from the saddle point A:(0,0), even though $\gp$ diverges as we evolve the system backward. This can also be verified if we take an initial condition in a small neighborhood of the fixed point A:(0,0) (like a perturbation from the fixed point) and evolve the system (\autoref{fig:thawdir} red dotted curve). The trajectory eventually reaches the fixed point C:(1,$\frac{\lambda^2}{3}$).


\section{Tracking Model}
The tracking model suggests that the equation of state parameter $w_\phi$ is decreasing gradually and have reached its current value $w_\phi\approx-0.95$ which is close to $-1$. It is called `tracking' because the energy density of the scalar field, for most of the evolution, tracks that of dark matter which gravitates in the known attractive fashion. The trackers have another very important significance, it is insensitive to initial scalar field energy density up to 100 orders of magnitude. Thus all the trajectories starting within this range approach to a common evolutionary track, known as the \textit{`tracker solution'}. In other words, there is no serious ``fine tuning'' issue in trackers.

Certain assumptions and conditions need to be satisfied for a tracking model. First of all, the change in $\gp$ with time is considered negligible. Another condition imposed on the potential is that $\Gamma \geq 1$ and $\Gamma$ is almost constant w.r.t. time, for the solutions to converge to the tracker solution. In this study we consider a near tracking case for which $\Gamma\approx1$, and Eq.\eqref{diff_eq_new} reduces to a 2D system with $\lambda=$ constant. 
The fixed points of this system are given in \autoref{tab:tab2}.
Point A, B and C are the same fixed points that appeared for the thawing model (\autoref{tab:tab1}). However, for the thawing models, C was a stable fixed point for all $\lambda$, whereas C is nonexistent in tracking models for $\lambda^2>6$.
It can be inferred that the system either evolves towards fixed points C or E depending on the value of $\lambda$. 
\begin{table}[!h]
\caption{Fixed points for a system with a tracking potential and their stability.}\label{tab:tab2}
\begin{center}
	\begin{tabular}{M{1cm}|M{5mm}|M{5mm}|M{1.5cm}|M{3.5cm}}
		Fixed point & $\op$ & $\gp$ & Existence & Stability\\
		\hline A. & 0 & 0 & $\forall$ $\lambda$ & saddle \TS\B\\
		\hline B. & 1 & 0 & $\forall$ $\lambda$ & saddle \TS\B \\
		\hline C. & 1 & $\frac{\lambda^2}{3}$ & $\lambda^2 < 6$ & \shortstack{stable if $\lambda^2 < 3$ \\ saddle if $3 \leq \lambda^2 < 6 $} \T\B \\
		\hline  D. & 1 & 2 & $\forall$ $\lambda$ & \T\shortstack{unstable node if $\lambda^2 <6 $ \\ saddle if $\lambda^2 > 6$} \B\\
		\hline E. & $\frac{3}{\lambda^2}$ & 1 & $\lambda^2 \geq 3$ & Stable \TS\B\\
		\hline F. & 0 & 2 & $\forall$ $\lambda$ & saddle \TS\B \\
		\hline 
	\end{tabular}
\end{center}
\end{table}

\begin{figure}
\centering
\includegraphics[width=1.0\linewidth]{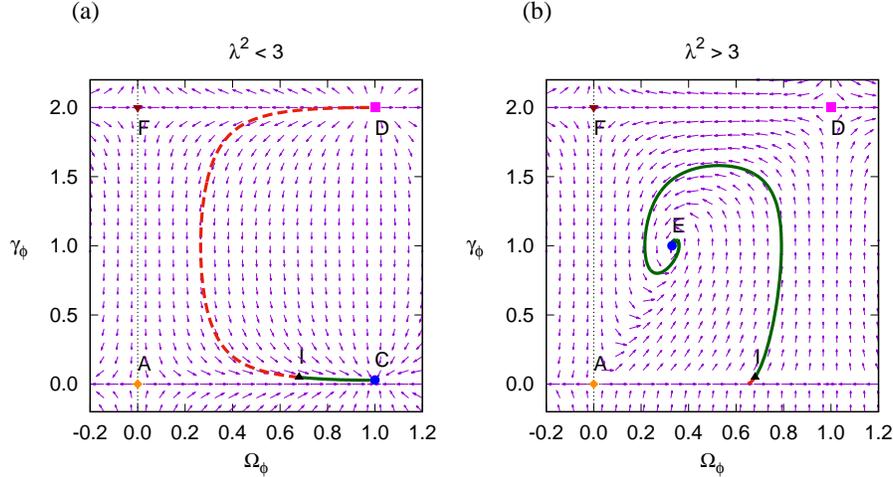}
\caption{Tracking model: Evolution of the universe in the $\gamma_\phi$ $-\hspace*{1mm}\Omega_\phi$ plane backward (red dashed line) and forward (green solid line) in time starting from the current value (I) of the scalar field energy density and effective EOS parameter.}
\label{fig:lam3}
\end{figure}

\autoref{fig:lam3} shows the direction field of the system for two different cases with $\lambda^2 < 3$ and $\lambda^2 > 3$. In the first case, a trajectory will approach C as it is a stable fixed point. For $\lambda^2<2$, the universe will continue in its phase of acceleration forever as this condition implies the final EOS parameter $(w_\phi)_{final}<-\frac{1}{3}$. For $2<\lambda^2<3$, stability requires the universe switching to a decelerated expansion phase in the future.

In the latter case ($\lambda^2 >3$) trajectories will converge to fixed point E whereas the fixed point C loses its stability, becoming a saddle or becomes nonexistent. In both the figures 4a and 4b, the evolution of the universe is shown  backward (red dashed line) and forward  (green solid line) in time with current observed values of $\op$ and $\gp$ taken as the initial condition (Point I in the figures).  
From \autoref{fig:lam3}(a), we see that the trajectory evolves back to point D, an unstable fixed point that can be thought of as the beginning of the universe. However, for $\lambda^2>3$ the system cannot be traced back to a fixed point (\autoref{fig:lam3}(b)).

\begin{figure}
\centering
\includegraphics[width=1.0\linewidth]{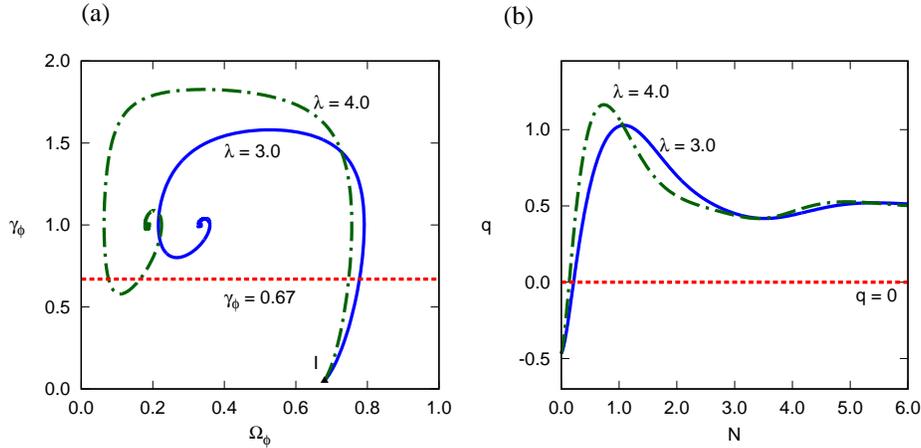}
\caption{(a) $\gp$ vs $\op$ plot for two different values of $\lambda$. For $\lambda$ = 4.0, the spiral crosses the line $\gp=\frac{2}{3}$.(b) $q$ vs $N$ plot for $\lambda$ = 3.0 and $\lambda$ = 4.0.}
\label{fig:fig_track}
\end{figure}

For $\lambda^2 > 3$, the stable fixed point is E=(3/$\lambda^2,1$). The eigenvalues of the Jacobian matrix for this fixed point are given as: $\Big\{\frac{1}{4}\Big(-3+\sqrt{\frac{24}{\lambda^2}-7}\Big), \frac{1}{4}\Big(-3-\sqrt{\frac{24}{\lambda^2}-7}\Big)\Big\}$. For $3<\lambda^2<\frac{24}{7}$, the eigenvalues show that the fixed point will be a stable node. For $\lambda^2>\frac{24}{7}$ the fixed point will be a stable spiral since the eigenvalues become complex with a negative real part. If the trajectory becomes a spiral then there is a possibility that it may cross the $\gp = 2/3$ line multiple times in the future, giving rise to multiple phases of acceleration and deceleration depending upon the value of $\lambda$. For higher values of $\lambda$, we have checked that the spiral can indeed cross the $\gp = 2/3$ line twice more (\autoref{fig:fig_track}a). In all such cases, the universe finally settles down to a decelerated phase. The accelerated phase appears to be a transient phenomenon (recall that $\gp < \frac{2}{3}$ is a necessary condition for the scalar field to act as a dark energy). \\

\begin{figure}
\centering
\includegraphics[width=1.0\linewidth]{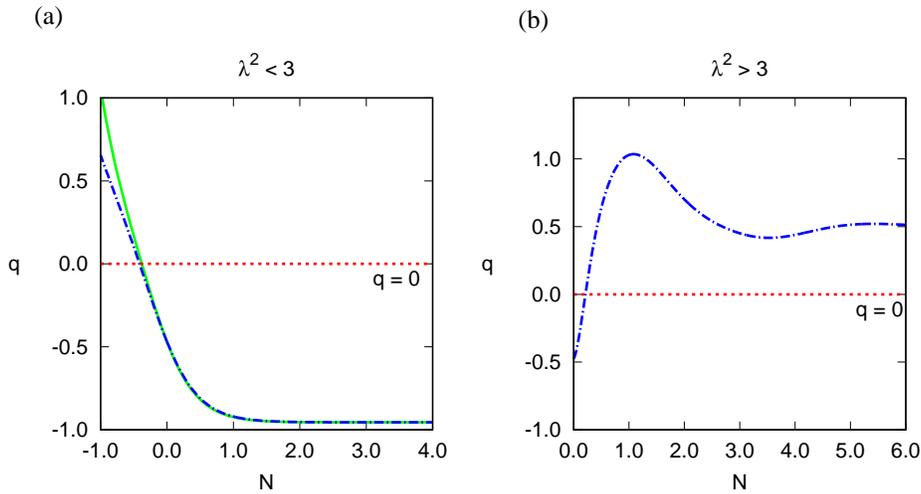}
\caption{q vs. N plot for both thawing (green solid line) and tracking model (blue dot-dashed line) for (a) $\lambda = 0.3$ (b) $\lambda = 3.0$.}
\label{fig:qcomp}
\end{figure}

The plot of $q$ vs. $N=\ln(a/a_0)$  shows that there is no phase of acceleration after the universe switches back to a decelerated expansion phase (\autoref{fig:fig_track}b). So, even though the \eosp of the scalar field becomes less than $2/3$, the \sed is low enough for the deceleration to continue. The switch to a decelerating phase from its current accelerating phase is a feature of the tracking model that happens around a redshift value $z \approx -0.15$ (or equivalently, $N \approx -0.16$). This feature is absent in the thawing case, where the ultimate fate of the universe is always an accelerated expansion phase. 
To compare the redshift value at the onset of the late time acceleration with the thawing model, we plot q vs. N for both models together for different regimes of $\lambda$ (\autoref{fig:qcomp}). One should note that for Figure 6b, the thawing case is irrelevant.

\begin{figure}
\centering
\includegraphics[width=0.7\linewidth]{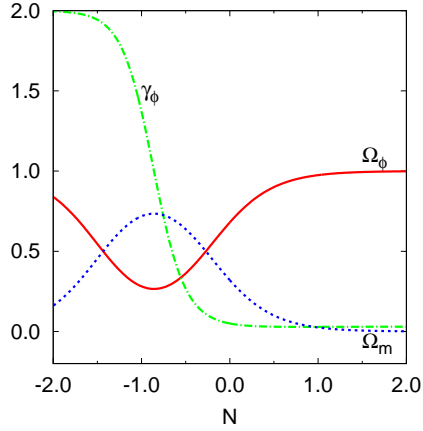}
\caption{The plot of $\op$, $\Omega_m$, and $\gp$ vs. N for tracking model ($\lambda^2<3$)}
\label{fig:tracktime}
\end{figure}

The variables $\op$, $\gp$, and $\Omega_m$ 
as a function of $N=\ln(a/a_0)$ remain bounded (for $\lambda^2<3$) (\autoref{fig:tracktime})
unlike the thawing model (\autoref{fig:thawtime}) where $\gamma_{\phi}$ has a divergence as we evolve the system backwards. 
This is also quite apparent from the direction field of the system shown in \autoref{fig:lam3}. 
The saddles at the fixed points A, B, F and unstable node D ensure that any trajectory that starts within this rectangular region will stay within this region. The absence of the fixed points D and F, in the case of a thawing model, allowed the trajectory to diverge.

Another point to note from \autoref{fig:tracktime} is that before the current phase of acceleration, there was a time when the \sed dominated over the matter energy density parameter. But the \eosp was sufficiently high ($\geq \frac{2}{3}$) to allow the universe to expand in a more sedate decelerated fashion. Therefore the deceleration parameter does not become negative until recently. This feature of a scalar field dominated but decelerated expansion is also absent in the thawing case.

\section{DISCUSSION}

In the present work, dark energy models that can be classified as either a thawing model or a tracking model are discussed in the context of dynamical stability. The field equations are written in terms of dimensionless variables in the form of an autonomous system of equations. The stability analysis shows that both the classes indeed have stable fixed points. No specific potential is considered in either case, only the relevant conditions for thawing and tracking are used in the analysis. \\

Indeed the tracking models have some features which the thawing models do not. For example, for the former the \eosp $\gamma_{\phi}$ for the scalar field is always well behaved but the latter has a divergence at some stage (\autoref{fig:tracktime} and \autoref{fig:thawtime}). The former also has more possibilities of stable fixed points namely $C$ and $E$ whereas the latter has only one in $C$. \\

The important thing to note is that for the stable fixed point E, in the case of tracking models, the stable value of \eosp is greater than unity (\autoref{tab:tab2}), and the accelerated phase, if there is any, is a transient phase. For ${\lambda}^2 < 2$, which easily satisfies the requirement of ${\lambda}^2 < 3$ for stability in this case, the fixed point C indeed has the possibility that an accelerated universe is a stable configuration. The thawing model, on the other hand, has one possibility of a stable fixed point, the relevant value of $\gamma_{\phi}$ is much less than $\frac{2}{3}$ (\autoref{tab:tab1}) as the value of $\lambda_0$ for a thawing case is already much less than unity as a condition of a slow-roll. The final stable configuration of thawing models is that of an accelerated universe.


\begin{thebibliography}{100}
\bibitem{perl} S. Perlmutter {\it et al}, Astrophys. J., {\bf 517}, 565 (1999).
\bibitem{schmidt}B.P. Schmidt {\it et al}, Astrophys. J.,{\bf 507}, 46,1998).
\bibitem{riess} A. Riess {\it et al}, Astron. J., {\bf 116}, 1009 (1998).
\bibitem{knopp} R. A. Knopp {\it et al}, Astrophys. J., {\bf 598}, 102 (2003).
\bibitem{planck} R. Adam {\it et al}, Astron. Astrophys., {\bf 594}, A8 (2016).
\bibitem{paddy} T. Padmanabhan, Phys. Rep., {\bf 380}, 235 (2003).
\bibitem{weinberg} S. Weinberg, Rev. of Mod. Phys., {\bf 61}, 1 (1989). 
\bibitem{sami} E.J. Copeland, M. Sami and S. Tsujikawa, Int. J. Mod. Phys. D, {\bf 15}, 1753 (2003).
\bibitem{varun} V. Sahni and A. Starobinsky, Int. J. Mod. Phys. D, {\bf 15}, 2015 (2006).
\bibitem{brax} P. Brax, Rept. Prog. Phys. {\bf 81}, 016902 (2018).
\bibitem{riess1} A. G. Riess, Astrophys. J., {\bf 560}, 49 (2001).
\bibitem{caldwell} R.R. Caldwell and E.V. Linder, Phys. Rev. Lett. {\bf 95}, 141301 (2005).
\bibitem{stein1} I. Zlatev, L. Wang and P. Steinhardt, Phys. Rev. Lett., {\bf 82}, 896 (1999).
\bibitem{stein2} P. Steinhardt, L. Wang and I. Zlatev, Phys. Rev. D, {\bf 59}, 123504 (1999).
\bibitem{johri} V. B. Johri, Phys. Rev. D, {\bf 63}, 103504 (2001).
\bibitem{anjan} R.J. Scherrer and A. A. Sen, Phys. D, {\bf 77}, 083515 (2008).
\bibitem{concord} L-M. Wang, R.R. Caldwell, J.P. Ostriker and P.J. Steinhardt, Astrophys. J., {\bf 538}, 17 (2000).
\bibitem{chiba} T. Chiba, Phys. Rev. D, {\bf 79}, 083517 (2009).
\bibitem{soma} S. Sen, A.A. Sen and M. Sami, Phys. Lett. B, {\bf 686}, 1 (2010).
\bibitem{chiba2} T. Chiba, A. DeFelice and S Tsujikawa, Phys. Rev. D, {\bf 87} (2013).
\bibitem{shruti} S. Thakur, A. Nautiyal, A.A. Sen and T. Seshadri, Mon. Not. R. Astron. Soc., {\bf 427}, 988 (2012).
\bibitem{glen94} P. Glendinning, {\it Stability, instability and chaos: an introduction to the theory of nonlinear differential equations}. Cambridge university press, 1994.
\bibitem{nandan} N. Roy and N. Banerjee, Gen. Rel. Grav., {\bf 46}, 1651 (2014).
\bibitem{ade et al} P. Ade {\it et al}, Astronomy and Astrophysics, {\bf 571}, A16 (2014).

\end{thebibliography}
\end{document}